# Seeing the unseen: observation of an anapole with dielectric nanoparticles

Andrey E. Miroshnichenko[1], Andrey B. Evlyukhin[2], Ye Feng Yu[3], Reuben M. Bakker[3], Arkadiy Chipouline[4], Arseniy I. Kuznetsov[3], Boris Luk'yanchuk[3], Boris N. Chichkov[2], Yuri S. Kivshar[1]

[1]Nonlinear Physics Centre, The Australian National University, Acton, ACT 2601, Australia

[2]Laser Zentrum Hannover e.V., Hollerithallee 8, D-30419 Hannover, Germany

[3]Data Storage Institute, A*STAR, 5 Engineering Drive 1, 117608, Singapore

[4] Institute of Applied Physics, Friedrich-Schiller-Universitaet Jena, 07743 Jena, Germany

*Nonradiating current configurations* attract attention of physicists for many years as possible models of stable atoms in the field theories [1, 2, 3]. One intriguing example of such a nonradiating source is known as "anapole" (which means "without poles" in Greek), and it was originally proposed by Yakov Zeldovich in nuclear physics [4]. Recently, an anapole was suggested as a model of elementary particles describing dark matter in the Universe [5]. Classically, an anapole mode can be viewed as a composition of electric and toroidal dipole moments [6], resulting in destructive interference of the radiation fields due to similarity of their far-field scattering patterns. Here we demonstrate experimentally that dielectric nanoparticles can exhibit a radiationless anapole mode in visible. We achieve the spectral overlap of the toroidal and electric dipole modes through a geometry tuning, and observe a highly pronounced dip in the far-field scattering accompanied by the specific near-field distribution associated with the anapole mode. The anapole physics provides a unique playground for the study of electromagnetic properties of nontrivial excitations of complex fields, reciprocity violation, and Aharonov-Bohm like phenomena at optical frequencies.

The term "anapole" was introduced in the physics of elementary particles [4]. The electrodynamics analog of a stationary anapole is the well-known toroid with a constant toroidal surface current [4]. This current distribution is also associated with a toroidal dipole moment pointing outward along the torus symmetry axis (see Fig. 1). The static magnetic field produced by a toroid is entirely concentrated within a coil in the form of a circulating magnetic current (see Fig. 1). It generates no field outside, but may possess a nonzero potential, which might lead to violation of the reciprocity theorem and Aharonov-Bohm like phenomena [7,8].

In the dynamic case, an oscillating toroidal dipole moment produces *nonzero electromagnetic radiation* with the pattern fully repeating the radiation pattern of an electric dipole moment but scaled by the factor of $\omega^2$, where $\omega$ is the angular frequency of light. For the oscillating surface current, its radiationless properties can be realized by adding the second electric dipole oscillating in out-of-phase with the toroid, resulting in complete destructive interference of their radiation due to similarity in the far-field scattering patterns, and such a radiationless nontrivial current configuration was named "anapole" [6]. However, this type of the radiation compensation is not complete: the compensated toroidal dipole moment is a part of the third-order multipole field expansion, and all higher order expansions remain radiative. Nontrivial nonradiative current configurations attracted continuous interest since the early days of electrodynamics; from the fundamental physics to applications of nonscattering objects [9,10]. Since ideal nonscattering objects do not exist [11], we understand by "radiationless" the compensation of the multipole expansion up to the third order; and the respective combination termed as anapole mode. The concept of toroidal modes attracted considerable attention in the field of metamaterials as a possible realization of radiationless objects [12-17]. The toroidal moment itself and respective effects (including toroidal metamaterials [17, 18]) have been studied theoretically [7,8,13,15,19,20]. Several experimental verifications in microwave [12, 14, 18] and optical [21] domains for toroidal moments confirmed the theory. Here, we demonstrate experimentally, for the first time to our knowledge, the existence of an anapole mode in optics in a simple structure of silicon nanoparticles. We observe the resonance suppression of the total far-field scattering along with evolution of the near-field mode around the nanodisk close to the wavelength of the anapole mode excitation.

In order to analyze the electromagnetic properties of the silicon nanoparticle theoretically, we employ multipole expansions in *Cartesian* and *Canonical (or Spherical) basis* [22]; each series can be unambiguously expressed one through the other [23]. For the sake of generality, assume a nontrivial current distribution $\mathbf{j}(\mathbf{r},t)$ producing an electromagnetic field $\mathbf{E}(\mathbf{r})$ [20, 24, 25]:

$$\mathbf{j}(\mathbf{r},t) = \sum_{l=0} \frac{(-1)^l}{l!} B_{i...k}^{(l)} \partial_i ... \partial_k \delta(\mathbf{r})$$

$$B_{i...k}^{(l)} = \int \mathbf{j}(\mathbf{r},t) \mathbf{r}_i ... \mathbf{r}_k d^3\mathbf{r}$$

(1)

Here $B_{i...k}^{(l)}$ is a tensor of rank $l$. From these tensors, various Cartesian multipoles can be obtained. For example, $B_i^{(1)} \sim d_i^{(1)}$ determines the electric dipole moment $d_i^{(1)}$, $B_{ij}^{(2)} \sim Q_{ij}^{(2)} + \mu_i^{(1)}$ consists of electric quadrupole $Q_{ij}^{(2)}$ (symmetric) and magnetic dipole $\mu_i^{(1)}$ (anti-symmetric) moments, and $B_{ijk}^{(3)} \sim O_{ijk}^{(3)} + \mu_{ij}^{(2)} + T_i^{(1)}$ gives rise to electric octupole $O_{ijk}^{(3)}$, magnetic quadrupole $\mu_{ij}^{(2)}$, and toroidal dipole moments $T_i^{(1)}$. Note here, that the toroidal dipole moment of the current appears in third order coefficients of expansion [24].

On the other hand, the radiation properties can by described by using the total scattering cross-section in *Canonical basis* can be written as a sum of intensities of spherical electric $a_E(l,m)$ and magnetic $a_M(l,m)$ scattering coefficients [23]:

$$C_{sca} = \frac{\pi}{k^2} \sum_{l=1}^{\infty} \sum_{m=-l}^{l} (2l+1)\left[\left|a_E(l,m)\right|^2 + \left|a_M(l,m)\right|^2\right] \quad (2)$$

which can be unambiguously determined by multipole Cartesian coefficients $B_{i...k}^{(l)}$. We are focusing on the situation when a spherical electric dipole mode is the dominant one. In this case the total scattering cross-section is determined solely by the electric dipole scattering coefficient $C_{sca} \propto \left|a_E(1,\pm 1)\right|^2$. Up to the first order expansion [23], the relation between Cartesian and spherical multipoles can be written as

$$a_E(1,\pm 1) = C_1\left[\pm B_x^{(1)} + iB_y^{(1)}\right] + 7C_3\left[\pm B_{xxx}^{(3)} + 2B_{xyy}^{(3)} + 2B_{xzz}^{(3)} - B_{yyx}^{(3)} - B_{zzx}^{(3)}\right] - i\left[B_{yyy}^{(3)} + 2B_{yxx}^{(3)} + 2B_{yzz}^{(3)} - B_{xxy}^{(3)} - B_{zzy}^{(3)}\right] \quad (3)$$

where $C_1 = -ik^3/(6\pi\varepsilon_0 E_0)$ and $C_3 = -ik^5/(210\pi\varepsilon_0 E_0)$. Thus, the total scattering can vanish if the electric dipole scattering coefficient becomes zero, $a_E(1,\pm 1) \approx 0$, provided all higher-order scattering amplitudes are also close to zero. In order to zero the spherical electric dipole, $a_E(1,\pm 1) = 0$, the first order (electric dipole) coefficients $B_i^{(1)}$ has to be compensated by the third order coefficients $B_{ijk}^{(3)}$, which contains the toroidal dipole moments $T_i^{(1)}$. This simple consideration creates the basis for understanding of physics of the anapole mode, namely mutual compensation of $B_i^{(1)}$ and $B_{ijk}^{(3)}$. The other two moments of the third order - octupole $O_{ijk}^{(3)}$ and magnetic quadrupole $\mu_{ij}^{(2)}$ are assumed to be negligible, which is true for the toroidal structure.

Figure 2 shows contributions of the spherical and Cartesian dipole moments to the total scattered field for a spherical electric dipole only excited inside a dielectric sphere. These results indicate that due to inhomogeneous field distributions inside optically large particles, it is necessary to introduce toroidal dipole moments to accurately describe the scattered field in terms of Cartesian multipoles. The total scattering cancellation is possible due to the fact that the radiation patterns of the electric and toroidal dipoles are equivalent.

A simple, realistic structure which supports the anapole mode is a silicon nanodisk of height 50 nm and diameter ranging from 200 nm to 400 nm (see Fig. 3). For this geometry, calculations show a strong dip in the far-field scattering spectrum (Fig. 3b), accompanied by a near-field enhancement inside and around the disk, indicating the presence of the combination of toroidal and dipole modes - anapole mode. Opposite circular displacement currents in the left and right hand sides of the disk (Fig. 3c) generate a circular magnetic moment distribution that is perpendicular to the disk surface (Fig. 3d). This provides a strong toroidal moment oriented parallel to the disk surface.

Excitation of this toroidal dipole and its destructive interference with the electric dipole is confirmed through Cartesian and Canonical dipole decompositions (see the 310 nm diameter disk in Fig. 3e&f). One advantage of such disks compared to other geometries, e.g. spheres, is that the other multipoles apart from the electric and toroidal dipoles are strongly suppressed, providing almost zero scattering at the anapole resonance (see also Extended Data Fig. 1 for a silicon disk with diameter of 200 nm). The anapole excitation in silicon disks is robust against incident angle and polarization (see also Extended Data Fig. 2).

To confirm these theoretical predictions and observe the anapole resonance in experiment, a series of silicon nanodisks with a height of 50nm and diameters ranging from 160 nm to 310 nm were fabricated on a quartz substrate using e-beam lithography. Far-field scattering spectra (Fig. 4a) were measured using single nanoparticle dark-field spectroscopy (see Methods). For disks with a diameter greater than 200 nm, a scattering dip appears around 550 nm; as the diameter increases, the dip redshifts and becomes more pronounced. The spectral position of this far-field scattering minimum is in good agreement with the theoretically predicted anapole resonance (see Fig. 3b for comparison). To show that the spectral dip in far-field scattering corresponds to the dark anapole mode excitation, the near-field distribution around the disks is mapped at multiple wavelengths using a near-field scanning optical microscope and a supercontinuum light source (see Methods). Representative experimental near-field maps for the 310 nm diameter disk at wavelengths around the anapole resonance are provided in Fig. 4b, compared directly with near-field maps of transverse electric and magnetic field components from simulations. Both electric and magnetic near-field components were collected by the aperture-type NSOM systems [26]. The experimental near-field maps show the evolution of the spectral response throughout the visible. As the wavelength approaches 620 nm, we begin to see the splitting of the central hot-spot into two separate spots. Close to the anapole resonance at 640 nm a new hot-spot appears in the middle of the disk and its intensity increases with wavelength. This mode evolution in the near-field experiments mimics theoretical simulations for both electric and magnetic fields. Just like the experiments, in the simulation, a new hot-spot appears in the center of the disk at wavelengths close to the anapole resonance. This is clearly observed for both electric and magnetic field components. Similar near-field behavior with an anapole resonance at 620 nm is observed for a disk with diameter of 285 nm (see Extended Data Fig. 3).

The anapole mode observed in the silicon nanodisk originates from mode interference. This radiationless excitation makes the nanodisk invisible in the far-field at the anapole mode wavelength. The anapole mode offers a new way to achieve an invisibility condition for lossless dielectric nanostructures based on the radiation scattering cancelation (originally proposed in [27] for metallic core-shell nanoparticles). Moreover, the relation between the electric and toroidal dipoles can be directly extended to magnetic moments and allow introducing a new class of magnetic toroidal and respective anapole moments. Recent observation of magnetic response of silicon nanoparticles [28-30] suggests that it could be an ideal platform for the demonstration of magnetic anapoles.

We also mention that the anapole mode is not only limited to the disk geometries but it can also be observed for spheres or other dielectric nanostructures where the electric dipole contribution vanishes due to a switch to a "higher order dipole" mode. For other geometries, however, the effect can be hidden by contributions of higher-order multipole modes (quadrupoles, octupoles, etc.). Similar effects can also be expected in metallic systems supporting "higher-order dipole" modes.

**Acknowledgements**

The work of AEM was supported by the Australian Research Council via Future Fellowship program (FT110100037). The authors at DSI were supported by DSI core funds. Fabrication, Scanning Electron Microscope Imaging and NSOM works were carried out in facilities provided by SnFPC@DSI (SERC Grant 092 160 0139). Zhen Ying Pan (DSI) is acknowledged for SEM imaging. Yi Zhou (DSI) is acknowledged for silicon film growth. Leonard Gonzaga (DSI), Yeow Teck Toh (DSI) and Doris Ng (DSI) are acknowledged for development of the silicon nanofabrication procedure. BNC acknowledges support from the Government of Russian Federation, Megagrant No. 14.B25.31.0019.

Figure 1: Illustration of an anapole excitation consisting of the superposition of electric and toroidal dipole moments. The toroidal dipole moment is associated with the circulating magnetic current **M** accompanied by electric poloidal current distribution. Since the symmetry of the radiation patterns of the electric **P** and toroidal **T** dipole modes are similar, they can destructively interfere leading to total scattering cancelation in the far-field with nonzero near-field excitation.

Figure 2: Contributions of the spherical and Cartesian dipole moments to the total scattered electric field for a spherical electric dipole only excited inside a dielectric sphere.
Calculated spherical electric dipole $\mathbf{P}_{sph}$ (black), Cartesian electric $\mathbf{P}_{car}$ (red) and toroidal $\mathbf{T}_{car}$ (green) dipole moments contributions [see Methods] to the scattered electric field for a lossless dielectric sphere as a function of radius for refractive index n=4 and wavelength 550nm. This figure demonstrates that for small particles both contributions of the spherical and Cartesian electric dipoles are identical and the toroidal moment is negligible. For larger sizes, the contribution of the toroidal dipole moments to the total scattered field has to be taken into account. The anapole excitation is associated with the vanishing of the spherical electric dipole $\mathbf{P}_{sph}=0$, when the Cartesian electric and toroidal dipoles cancel each other $\mathbf{P}_{car}=-\frac{ik}{c}\mathbf{T}_{car}$. Another interesting point is that Cartesian electric dipole contributions can also vanish $\mathbf{P}_{car}=0$ for certain dimensions, which can be associated with the pure toroidal dipole excitation.

Figure 3: Numerical near- and far-field properties for a single silicon nanodisk under normal incidence. **a,** Electric field on top of the nanodisk at the surface center for thickness h=50nm for various diameter. **b,** Total scattering cross-section spectra of the silicon nanodisks with similar parameters . The resonant scattering suppression is accompanied by the electric field enhancement in the nanodisk. **c-d**, Electric and magnetic scattered near-field distribution at the anapole resonance corresponding to the mimimum of the far-field scattering. **e-f**, Canonical and Cartesian multipole decomposition of the scattering spectra for diameter 310nm. In spherical harmonics only the electric dipole is dominant which becomes resonantly suppressed at 650nm wavelength. In the Cartesian basis there are two leading contributions from electric $\mathbf{P}_{car}$ and torodial $\mathbf{T}_{car}$ dipoles, which are out of phase and, thus, compensate each other in the far-field.

Figure 4: Experimental demonstration of the anapole excitation in a single silicon nanodisk. **a,** Experimental dark field scattering spectra of silicon nanodisks with a height of 50 nm and diameter ranging from 160 nm to 310 nm. The baseline for each spectrum has an offset step of 500 A.U. for viewing convenience (inset shows tilted (52°) SEM images of the two largest disks). **b**, Near-field enhancement around the silicon nanodisk with height of 50 nm and diameter of 310 nm; the top row shows experimental NSOM measurements while the middle and bottom rows show calculated transversal electric and magnetic near-field respectively on top of the disk, 10 nm above disk surface. White dashed lines in the experimental images indicate the disk position. Polarization of the excitation light is shown in the figure.

**Fig.1**

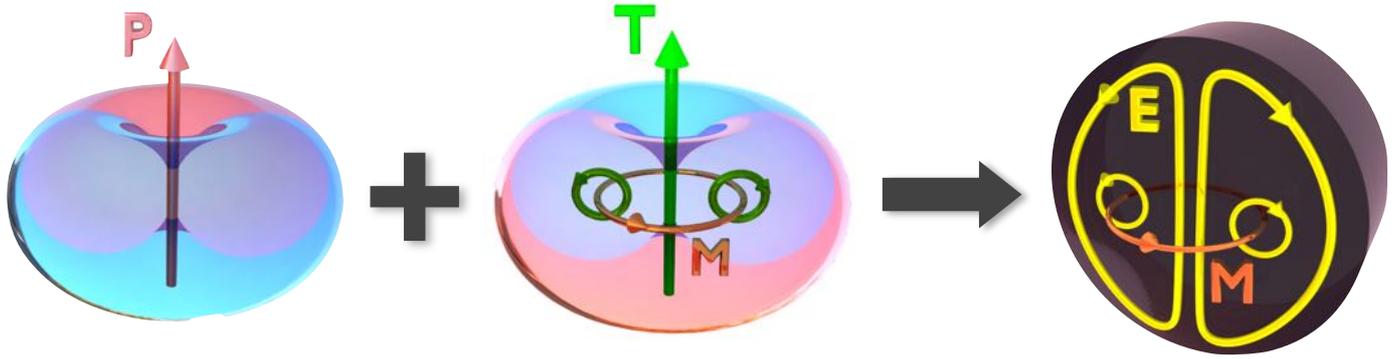

Electric dipole     Toroidal dipole     Anapole

**Fig.2**

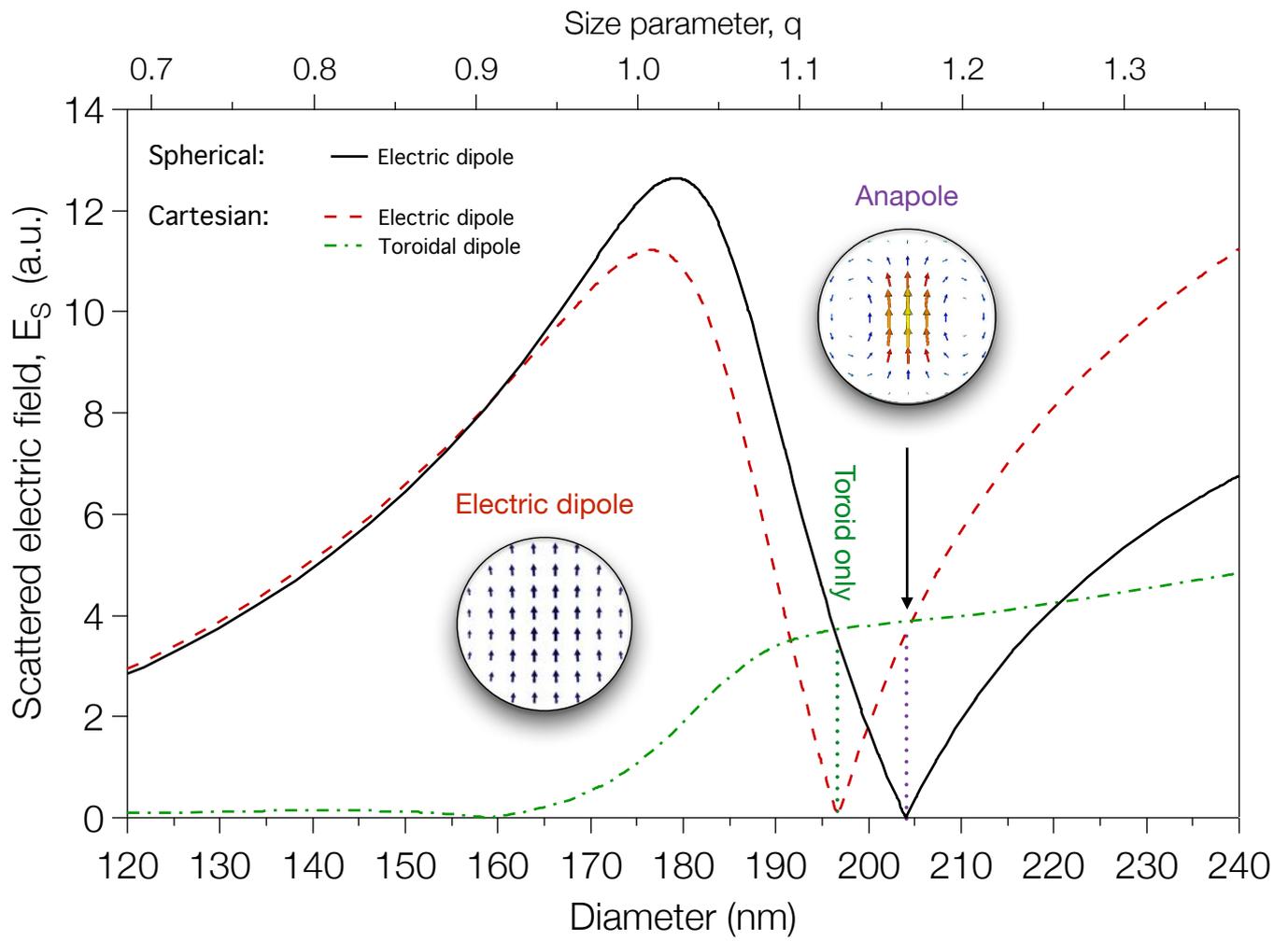

**Fig.3**

### Electric field on top

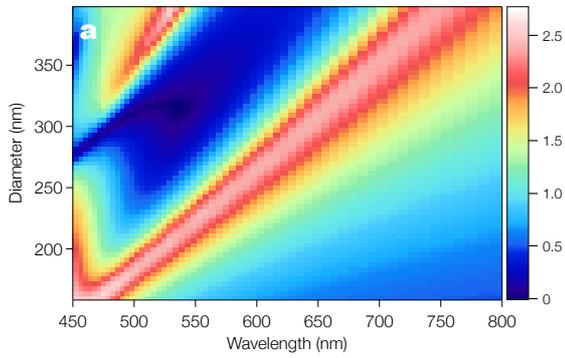

### Anapole field distribution

Electric

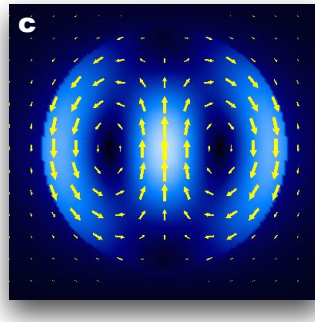

Magnetic

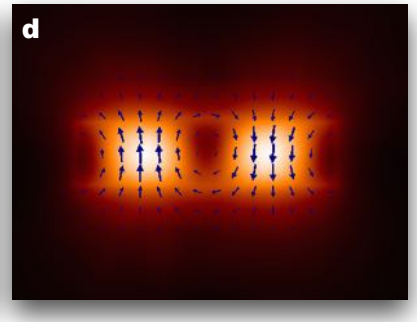

### Scattering cross-section

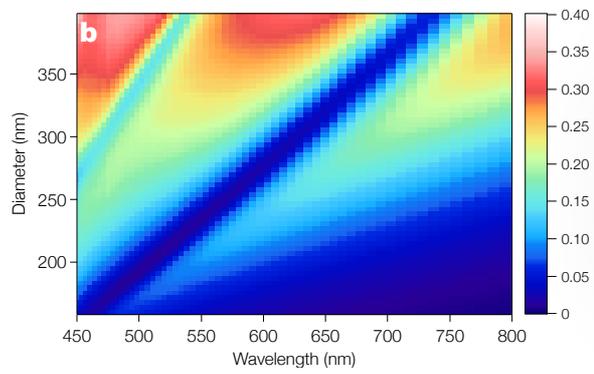

### Multipoles decomposition, D=310nm

Spherical

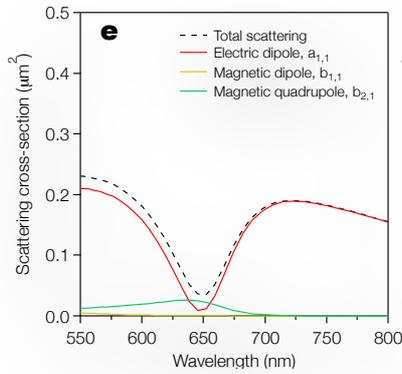

Cartesian

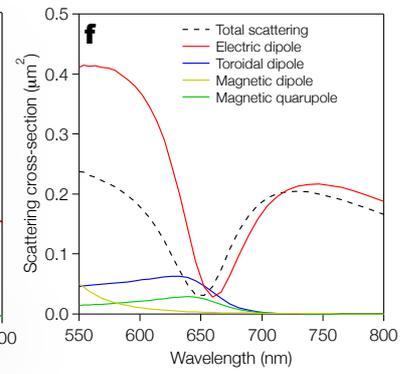

**Fig.4**

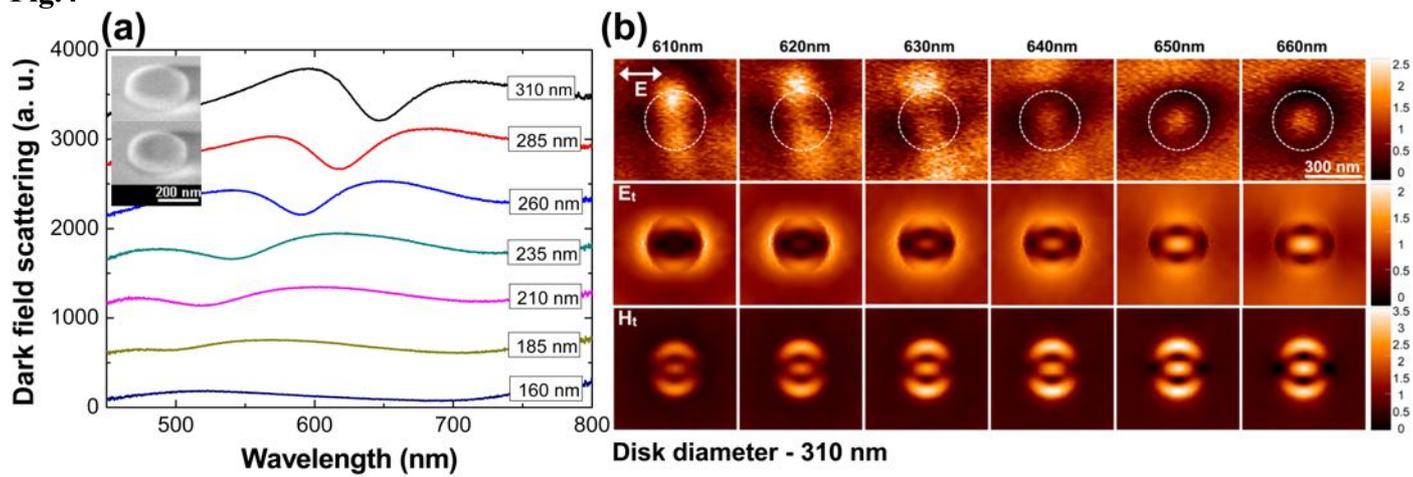

**Methods**.

*Fabrication*

Silicon nanodisks with various diameters were fabricated on quartz substrates by electron beam lithography (Elionix 100KV EBL system). A 50 nm silicon film was grown on a quartz substrate using chemical vapor deposition (ICP-CVD, Oxford Instruments). A thin (<60 nm) layer of negative resist (HSQ) was coated on the sample. After electron beam exposure and development, reactive ion etching (ICP, Oxford Instruments) was used to transfer the pattern into the silicon film. The result is silicon disks with a thin (<10 nm) cap of residual resist on top of a quartz substrate.

*Far-field spectroscopy*

Spectral analysis was performed using a single-nanoparticle spectroscopy setup with a dark-field geometry (see Ref. [30] for details). The sample was irradiated by a halogen lamp source at an angle of 58.5 degrees. According to our simulations the spectral position of the anapole resonance practically does not change for variations of the incidence angle in the range from 0° to 60° (see Extended Data Fig. 3). Scattering by the nanostructure is collected from the top into a solid angle corresponding to the microscope objective lens with 0.55 NA. The collected scattering spectra were normalized to the halogen lamp spectrum measured in bright-field reflection geometry.

*Near-field*

The sample was characterized in the near-field using a Near-Field Scanning Optical Microscope (Nanonics). The sample was illuminated from the far-field and the near-field was collected through a subwavelength aperture (50 nm aperture of a tapered fiber, coated with Chrome and Gold) in a transmission configuration. The light source used is a supercontinuum source (SuperK Power, NKT Photonics). Specific wavelengths were selected using a variable bandpass filter (SuperK Varia, NKT Photonics). Photons were counted with Avalanche Photo Diodes (Excelitas Technologies). Scans on individual particles were performed with various wavelengths over a 2x2 micrometer area with a pixel size of 8nm. The near-field maps are normalized, taking the background as unity.

*Comparison of Spherical and Cartesian dipoles contributions to the scattered field*

For further insight into the origin of the scattering cancelations, we can consider a simplified situation of light scattering by a hypothetical spherical particle where we take scattering contributions only from the spherical electric dipole mode excited inside the sphere [see Fig.2]. In this case, the total scattered electric field can be expressed independently in two bases. In the *Canonical (Sphereical)* basis there is only one contribution coming from the spherical electric dipole

$$\mathbf{E}_{sca} = \frac{k^2}{4\pi\varepsilon_0} \mathbf{n} \times \mathbf{P}_{sph} \times \mathbf{n}$$

with

$$\mathbf{P}_{sph} = \frac{6i\pi\varepsilon_0}{k^3} a_E(1,1)$$

In the *Cartesian* basis there are two leading contributions from the electric and toroidal dipole moments [24,25]

$$\mathbf{E}_{sca} = \frac{k^2}{4\pi\varepsilon_0}\left\{\mathbf{n}\times\mathbf{P}_{car}\times\mathbf{n} + \frac{ik}{c}\mathbf{n}\times\mathbf{T}_{car}\times\mathbf{n}\right\}$$

where

$$\mathbf{P}_{car} = \frac{1}{i\omega}\int\mathbf{J}\,d\mathbf{r}$$

and

$$\mathbf{T}_{car} = \frac{1}{10}\int\left[(\mathbf{r}\cdot\mathbf{J})\mathbf{r} - 2r^2\mathbf{J}\right]d\mathbf{r}$$

with

$$\mathbf{J} = -i\omega\varepsilon_0\left[\varepsilon_{sph} - 1\right]\mathbf{E}_{sph},$$

where the internal field can be analytically expressed as follows

$$\mathbf{E}_{sph} \propto d_1 N_{e11}^{(1)} = d_1\left(2\frac{j_1(\rho)}{\rho}\cos(\phi)P_1^1(\cos\theta)\,\hat{\mathbf{e}}_r + \cos(\phi)\frac{dP_1^1(\cos\theta)}{d\theta}\frac{1}{\rho}\frac{d}{d\rho}[\rho j_1(\rho)]\,\hat{\mathbf{e}}_\theta - \sin(\phi)\frac{P_1^1(\cos\theta)}{\sin\theta}\frac{1}{\rho}\frac{d}{d\rho}[\rho j_1(\rho)]\,\hat{\mathbf{e}}_\phi\right)$$

with

$$d_1 = \frac{n_{sph}j_1(\rho)[\rho h_1^{(1)}(\rho)]' - n_{Si}h_1^{(1)}(\rho)[\rho j_1(\rho)]'}{n_{sph}^2 j_1(n_{sph}\rho)[\rho h_1^{(1)}(\rho)]' - h_1^{(1)}(\rho)[n_{sph}\rho j_1(n_{sph}\rho)]'}$$

and $\rho = kR_{sph}$ being size parameter, $n_{sph}$ is refractive index of the particle, $j_l(x)$ is spherical Bessel function, $P_1^1(\cos\theta)$ is associated Legendre polynomial [31].

*Multipole expansion in Spherical harmonics*

Electromagnetic properties of silicon nanodisks were numerically studied by using CST Microwave Studio. In the Canonical basis we perform a multipole expansion of the scattered field of the nanodisk into vector spherical harmonics, which form a complete and orthogonal basis allowing the unique expantion of any vectorial field. To calculate electric $a_E(l,m)$ and magnetic $a_M(l,m)$ spherical multipole coefficients, we project the scattered electric field $\mathbf{E}_{sca}$ on a spherical surface, enclosing the silicon nanodisk centered at the symmetric point, onto vector spherical harmonics based on the following relations [32]:

$$a_E(l,m) = \frac{(-i)^{l+1}kR}{h_l^{(1)}(kR)E_0\sqrt{\pi(2l+1)(l+1)l}}\int_0^{2\pi}\int_0^\pi Y_{lm}^*(\theta,\phi)\hat{\mathbf{r}}\cdot\mathbf{E}_{sca}(\mathbf{r})\sin\theta\,d\theta\,d\phi$$

$$a_M(l,m) = \frac{(-i)^l kR}{h_l^{(1)}(kR)E_0\sqrt{\pi(2l+1)}}\int_0^{2\pi}\int_0^\pi \mathbf{X}_{lm}^*(\theta,\phi)\cdot\mathbf{E}_{sca}(\mathbf{r})\sin\theta\,d\theta\,d\phi$$

where $R$ is the radius of the enclosing sphere, $k$ is the wavevector, $h_l^{(1)}$ is the Hankel function with the asymptotic of the outgoing spherical wave, $E_0$ is the amplitude of the incident wave, $Y_{lm}$ and $\mathbf{X}_{lm}$ are scalar and vector spherical harmonics. Due to azimuthal symmetry of the silicon nanodisk the amplitude of the scattering coefficients with opposite $m$ indices are identical, $|a_{E,M}(l,m)| = |a_{E,M}(l,m)|$ [31].

For the anapole excitation discussed in this Letter, the dominant contribution to the scattering is given by the spherical electric dipole $C_{sca} \propto |a_E(1,1)|^2$.

*Cartesian multipoles and discrete dipole approximation*

Additional multipole analysis of scattering were performed using the decomposed discrete dipole approximation (DDDA) [25,33]. In this approach the scattering object is replaced by a cubic lattice of electric dipoles with the polarizability $\alpha_j$. The total number of dipoles is $N$. The corresponding dipole moment $\boldsymbol{p}_j$ induced in each lattice point $j$ (with the radius-vector $\boldsymbol{r}_j$) is found by solving the coupled-dipole equations. The Cartesian multipole moments of the scattering object (electric dipole moment $\mathbf{p}$, electric quadrupole moment $\hat{Q}(\boldsymbol{r}_0)$, magnetic dipole moment $\mathbf{m}(\boldsymbol{r}_0)$, toroidal dipole moment $\mathbf{T}(\boldsymbol{r}_0)$, and magnetic quadrupole moment $\widehat{M}(\boldsymbol{r}_0)$ located at a point $\boldsymbol{r}_0$) are simply calculated from the space distribution of $\boldsymbol{p}_j$

$$\mathbf{p} = \sum_{j=1}^{N} \boldsymbol{p}_j; \qquad \hat{Q}(\boldsymbol{r}_0) = \sum_{j=1}^{N} \hat{Q}^j(\boldsymbol{r}_0); \qquad \mathbf{m}(\boldsymbol{r}_0) = \sum_{j=1}^{N} \boldsymbol{m}_j(\boldsymbol{r}_0);$$

$$\widehat{M}(\boldsymbol{r}_0) = \sum_{j=1}^{N} M^j(\boldsymbol{r}_0); \qquad \mathbf{T}(\boldsymbol{r}_0) = \sum_{j=1}^{N} \boldsymbol{T}_j(\boldsymbol{r}_0);$$

where the corresponding Cartesian multipole moments associated with the single electric dipole $\boldsymbol{p}_j$, are under the summation symbols. Analytical expressions for these multipole moments are presented as

$$\hat{Q}^j(\boldsymbol{r}_0) = 3\left((\boldsymbol{r}_0 - \boldsymbol{r}_j) \otimes \boldsymbol{p}_j + \boldsymbol{p}_j \otimes (\boldsymbol{r}_0 - \boldsymbol{r}_j)\right), \qquad \boldsymbol{m}_j(\boldsymbol{r}_0) = -\frac{i\omega}{2}[(\boldsymbol{r}_0 - \boldsymbol{r}_j) \times \boldsymbol{p}_j],$$

$$\widehat{M}^j(\boldsymbol{r}_0) = -\frac{i\omega}{3}\{[(\boldsymbol{r}_0 - \boldsymbol{r}_j) \times \boldsymbol{p}_j] \otimes (\boldsymbol{r}_0 - \boldsymbol{r}_j) + (\boldsymbol{r}_0 - \boldsymbol{r}_j) \otimes [(\boldsymbol{r}_0 - \boldsymbol{r}_j) \times \boldsymbol{p}_j]\},$$

$$\boldsymbol{T}_j(\boldsymbol{r}_0) = -\frac{i\omega}{10}\left[\left((\boldsymbol{r}_0 - \boldsymbol{r}_j) \cdot \boldsymbol{p}_j\right)(\boldsymbol{r}_0 - \boldsymbol{r}_j) - 2(\boldsymbol{r}_0 - \boldsymbol{r}_j)^2 \boldsymbol{p}_j\right],$$

where $\omega$ is the angular frequency of scattered light.

The scattered electric field in the far field zone is presented by [25,33]

$$\mathbf{E}_S(r) = \frac{k_0^2 e^{ik_d(r-n r_0)}}{4\pi\varepsilon_0 r}\left([\mathbf{n}\times[\mathbf{p}\times\mathbf{n}]] + \frac{ik_d}{6}\left[\mathbf{n}\times[\mathbf{n}\times\hat{Q}(\boldsymbol{r}_0)\mathbf{n}]\right] + \frac{1}{v_d}[\mathbf{m}(\boldsymbol{r}_0)\times\mathbf{n}] + \frac{ik_d}{2v_d}\left[\mathbf{n}\times\widehat{M}(\boldsymbol{r}_0)\mathbf{n}\right] \right.$$
$$\left. + \frac{ik_d}{v_d}[\mathbf{n}\times[\mathbf{T}(\boldsymbol{r}_0)\times\mathbf{n}]]\right),$$

where $k_0$, $k_d$ and $v_d$ are the wavenumber in vacuum, and the wavenumber and light phase velocity in the medium surrounding the scatterer, respectively. $\varepsilon_0$ is the vacuum dielectric constant, $\mathbf{n}$ is the unit vector directed along the radius vector $\boldsymbol{r}$. Using the scattered electric field we calculated the total and angular distributions of scattered powers and to estimate contributions of every multipole moment [33]. Note that here $\widehat{M}(\boldsymbol{r}_0)$ is symmetrical and traceless [25]. Contribution of the electric octupole moment in our case is significantly small, so we excluded it from the consideration.

Extended Data Figure 1: Spherical multipole decomposition of the scattering spectra of a silicon nanodisk with height of 50 nm and diameter 200 nm. In this case, at the anapole excitation the dominant contribution is the electric dipole only, while all other modes are significantly suppressed.

Extended Data Figure 2: Numerical results for light scattering by silicon nanodisks at oblique incidence.
(a) Schematic of the simulation setup; (b, c) the total scattering cross-sections for TE and TM polarizations; (d,f) Cartesian electric and (e,g) toroidal dipoles contributions calculated by DDDA method for a silicon nanodisk with diameter 270 nm. These results demonstrate independence from incident angle and polarization.

Extended Data Figure 3: Near-field distribution around the silicon nanodisk with a height of 50 nm and diameter of 285 nm. The top row shows experimental NSOM measurements while the middle and bottom rows show calculated transversal electric and magnetic near-field respectively on top of the disk. White dashed lines in the experimental images indicate the disk position.

Extended Data Figure 1

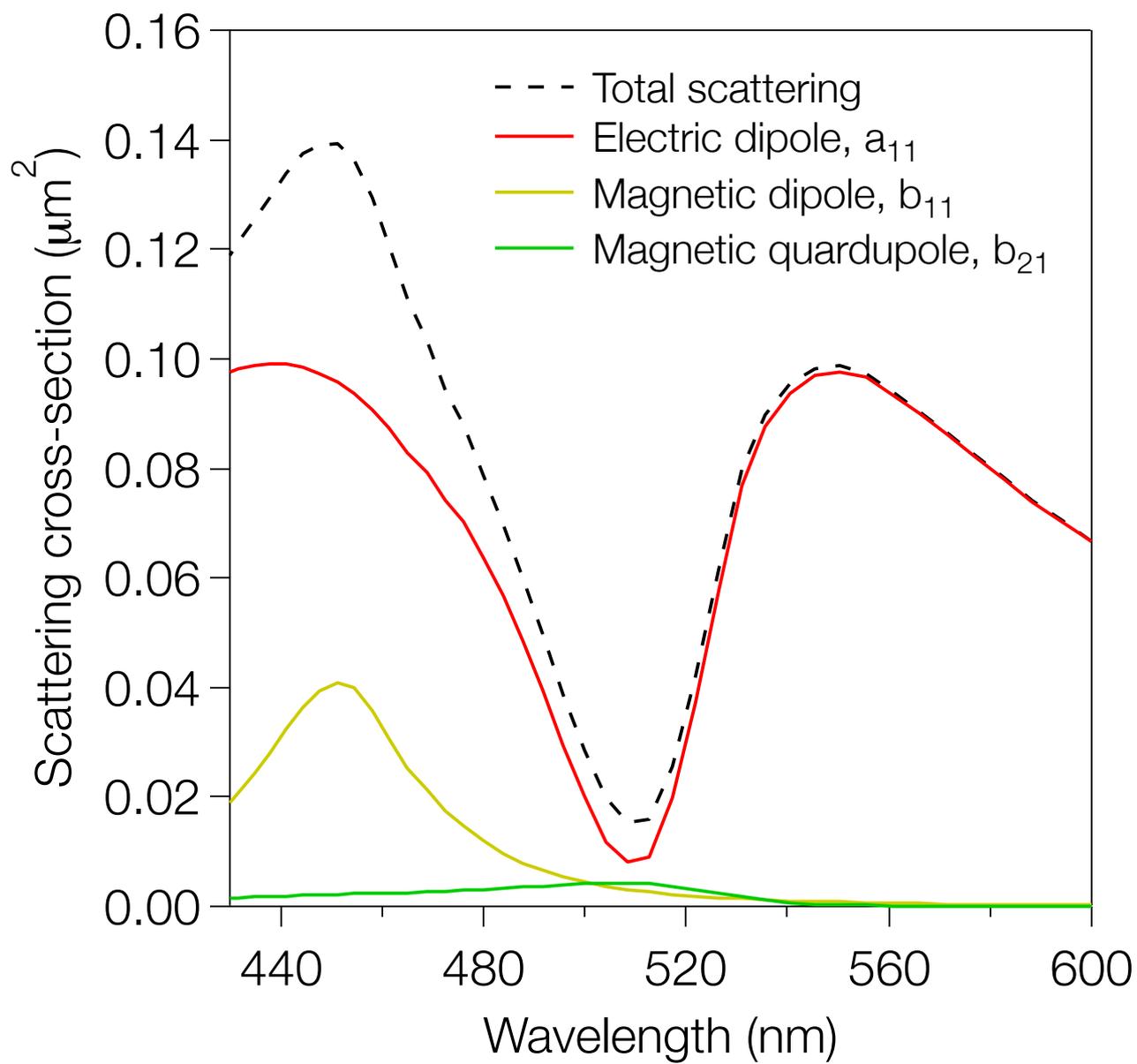

Extended Data Figure 2

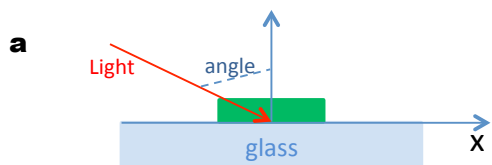
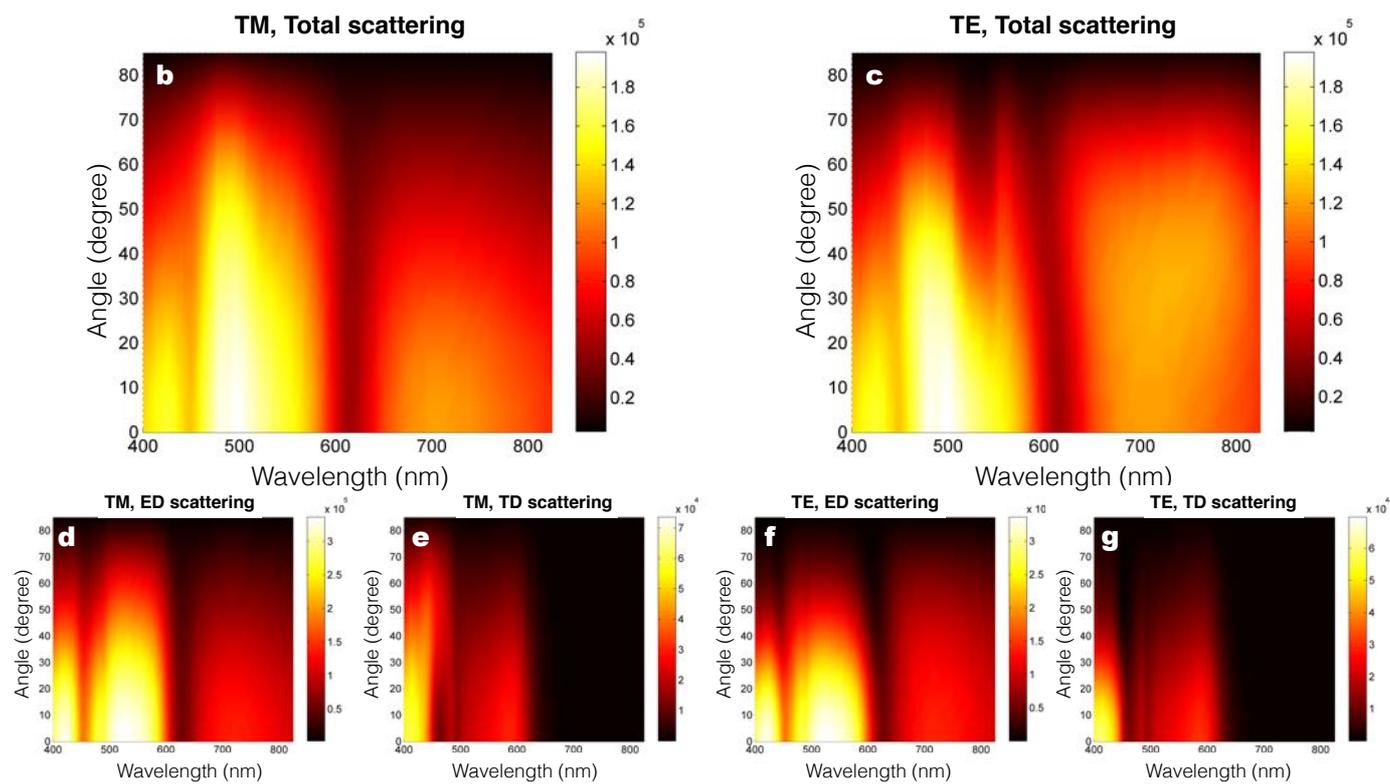

Extended Data Figure 3

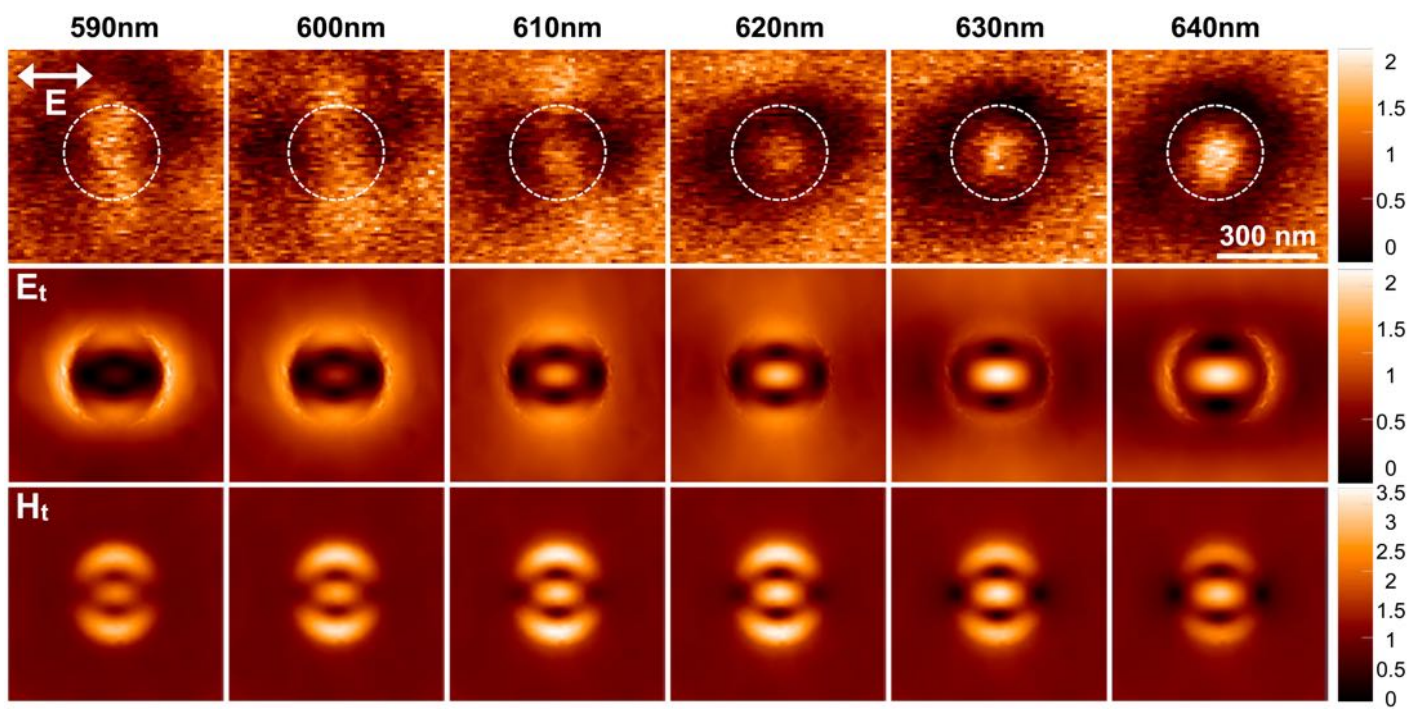